\documentclass[twocolumn,aps,footinbib,superscriptaddress]{revtex4-1}
\usepackage{amsmath,amssymb,bm,graphicx,hyperref}
\usepackage{latexsym}
\usepackage{bm}
\usepackage{braket}
\usepackage{amsmath}
\usepackage{amssymb}
\usepackage{color}
\usepackage[dvipsnames]{xcolor}
\usepackage{graphics}
\usepackage{graphicx}
\usepackage{eepic}
\usepackage{pstcol}
\usepackage{pstricks}%
\usepackage{pst-eps}%
\usepackage{pst-plot}%
\usepackage{pst-node}%

\renewcommand{\i}{{\rm i}}

\begin{document}
\title{Exact Plaquette-Ordered Ground States with Exact Edge States\\
of the Generalized Hubbard Model in Corner Sharing Lattices}
 \author{Naoto Nakatsuji}
\affiliation{
Department of Physics, Ehime University Bunkyo-cho 2-5, Matsuyama, Ehime
790-8577, Japan}
 \author{Satoshi Nishimoto}
\affiliation{
Department of Physics, Technical University Dresden, 01069 Dresden,
Germany}
\affiliation{
Institute for Theoretical Solid State Physics, IFW Dresden, 01171
Dresden, Germany}
 \author{Masaaki Nakamura}
\affiliation{
Department of Physics, Ehime University Bunkyo-cho 2-5, Matsuyama, Ehime
790-8577, Japan}
\date{\today}
\begin{abstract}
 We discuss exact plaquette-ordered ground states of the generalized
 Hubbard model based on the projection operator method for several
 corner sharing lattices: Kagome, checkerboard, and pyrochlore lattices.
 The obtained exact ground states are interpreted as N\'eel ordered
 states on the plaquette-located electrons.  We demonstrate that these
 models also have exact edge states.
 We also calculate the entanglement entropy
 exactly in these systems.
\end{abstract}

\maketitle

\section{Introduction}

In condensed matter physics, a major theoretical aim is to build an
effective model - as simple as possible - which captures the essence of
observed physical phenomenon. A prime example to describe quantum
mechanical motion of electrons in a solid is the Hubbard
model~\cite{Hubbard,Kanamori,Gutzwiller}.  The original Hubbard model
contains only two approximated components: (1) Electron transfer $t$ as
overlap integral between neighboring atoms and (2) on-site repulsion $U$
as intra-atomic Coulomb interaction on the assumption of Wannier basis;
and, the other site-off-diagonal interactions are neglected. Despite its
simple Hamiltonian, a variety of interesting phenomena such as
metal-insulator transition~\cite{Mott1990},
ferromagnetism~\cite{Tasaki2008}, antiferromagnetism~\cite{Fazekas1999},
Tomonaga-Luttinger liquid~\cite{Solyom1979}, and
superconductivity~\cite{Anderson1987}, etc. can be well explained.
Especially, experimental realization of the {\it ideal} Hubbard model
using ultracold fermions in optical lattices has been a hot topic in
recent years~\cite{Jaksch2005,Mazurenko2017}.  Nevertheless, it would be
also true that the original Hubbard model is too much oversimplified to
faithfully describe actual existing solids. In fact, the Hubbard model
with the site-off-diagonal interactions (referred as `generalized
Hubbard model') has been fairly studied in the context of magnetism and
superconductivity~\cite{Hirsch1,Hirsch2,Hirsch3,Hirsch4,Campbell-G-L1988,
Campbell-G-L1990,Simon-A,Ovchinnikov1993,Strack-V1993,Strack-V1994,
Arrachea-A,Boer-K-S,Boer-S,Montorsi-C,Kollar-S-V,Anfossi-D-M,
Arrachea-A-G,Millan-P-W,Dobry-A}.

\begin{figure}[h]
 \includegraphics[width=80mm]{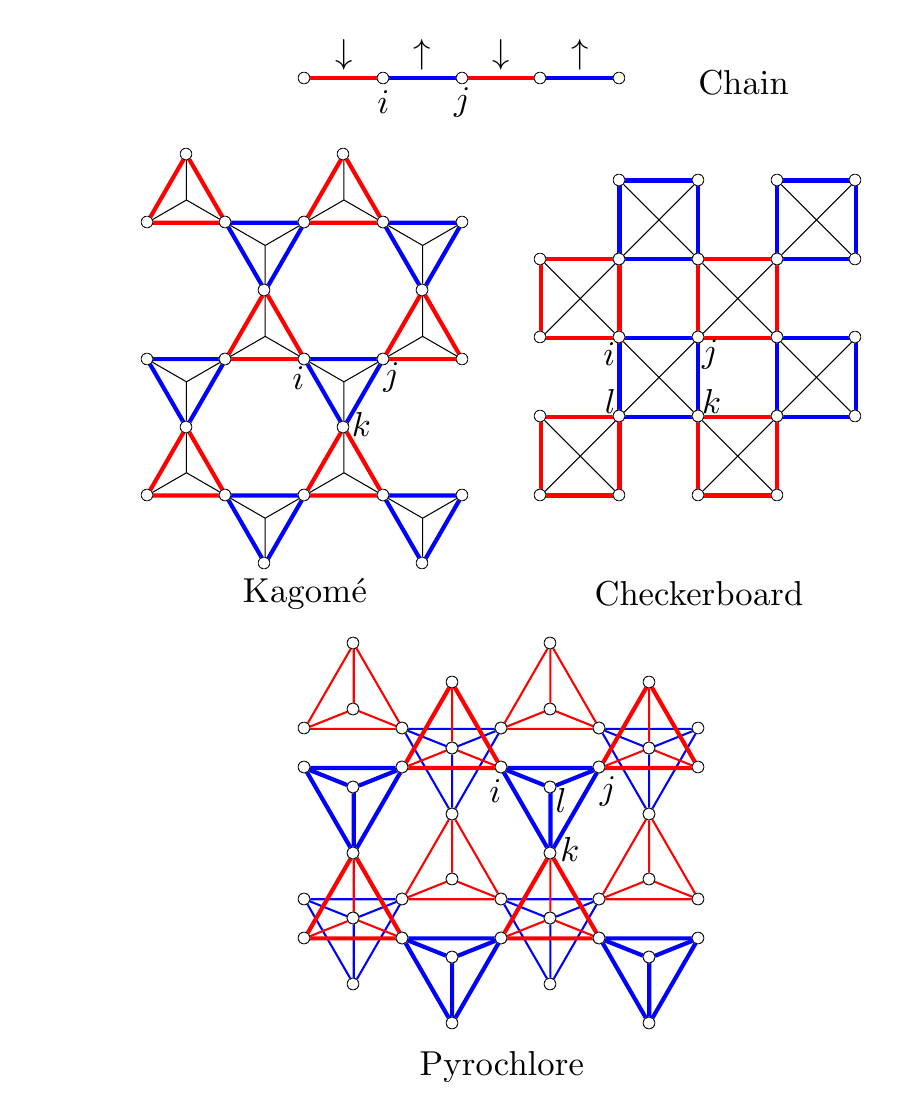}
 \caption{Examples of lattice structure where generalized Hubbard models
 with exact plaquette-ordered ground states can be constructed: the 1D
 chain, the Kagom\'e, the Checkerboard and the Pyrochlore lattices. The
 solid and the dashed plaquettes denote those belong to the groups
 ${\cal A}$ and ${\cal B}$, respectively.}\label{fig:lattices}
\end{figure}

The conventional Hubbard model has so far been analytically solved only
for the one-dimensional (1D) case. On the other hand, the generalized
Hubbard model may be more flexible at obtaining exact ground states
although with some restrictions on the interaction
parameters~\cite{Strack-V1993,Strack-V1994,Arrachea-A,Boer-K-S,
Boer-S,Montorsi-C,Kollar-S-V,Arrachea-A-G,Anfossi-D-M}. The basic idea
is that the ground state energy is the lower bound found by
diagonalizing the local Hamiltonian~\cite{Boer-S}. Based on this idea,
in order to obtain exact ground states of arbitrary dimensional
generalized Hubbard model, a more sophisticated treatment, using the
projection operator method~\cite{Majumder-G,Affleck-K-L-T,AKLT2} for
multi-component systems, was proposed~\cite{Itoh}. Adopting this method
in a simple 1D system, three kinds of ground states were recognized in a
wide range of parameter region; namely, ``bond N\'eel'' (BN),
ferromagnetic (FM), and phase separated
states~\cite{Itoh-N-M,Nakamura-I,Nakamura-O-I}. The BN state is regarded
as a N\'eel ordered state of bond-located spins. The concept of BN state
in one dimension can be extended to higher dimensional systems by
introducing multiplet states in corner sharing
lattices~\cite{Nakamura-I,Nakamura-N}.  For example, illustrated in
Fig.~\ref{fig:lattices}, the Kagom\'e (Checkerboard) lattice can be
covered by two colored multiplets alternatively, where each of the
multiplets consists of three (four) electrons with the same spin and the
spins belonging to different colored multiplets are antiparallel. These
staggered states can be regarded as antiferromagnetism on a honeycomb
(square) lattice. We call this state ``plaquette N\'eel (PN)'' state as
an extension of the BN state, Such the PN state is also realized in
three dimensional systems like the Pyrochlore lattice.
These plaquette ordered states are also important in discussions related
with Berry phases and higher order topological
states\cite{Hatsugai-M,Araki-M-H}.

In this paper, we study the generalized Hubbard model on corner sharing
lattices: Kagom\'e, Checkerboard, and Pyrochlore lattices. Based on the
projection operator method, various kinds of exact plaquette-ordered
ground states are found at commensurate fillings. The obtained ground
states are interpreted as the PN ones.  We also suggest that exact edge
states are constructed in the presence of free boundary. Furthermore,
the entanglement entropy in the PN states is calculated.

This paper is organized as follows: In Section~\ref{sec:method}, we
explain the projection operator method to construct Hamiltonians with
exact ground states in multicomponent systems.  In
Section~\ref{sec:exact}, we review the application of this method to the
1D and Kagom\'e systems. The estimation of exact edge states is also
demonstrated.  In Section~\ref{sec:checker}, we perform the projection
operator analysis for the Checkerboard lattice and obtain the exact PN
states at 1/4, 1/2, and 3/4 fillings.  In Section~\ref{sec:pyro}, like
in Section~\ref{sec:checker} we obtain the exact PN states for the
Pyrochlore lattice at 1/4 and 3/4 fillings. In Section~\ref{sec:EE}, we
calculate the entanglement entropy for the PN states of Checkerboard and
Pyrochlore systems. Finally, we give summary and discussion of the
results in Section~\ref{sec:summary}.

\section{Parent Hamiltonian}
\label{sec:method}

The method to construct a Hamiltonian with an exact ground state is the
following way~\cite{Itoh}. First, we consider a Hamiltonian given by a
sum of products of projection operators
\begin{align}
 {\cal H}&=\sum_{\alpha} h_{\alpha},\quad
  h_{\alpha}=\sum_{\mu,\nu}\lambda_{\mu\nu}
  R^{(\mu)}_{\alpha\uparrow}R^{(\nu)}_{\alpha\downarrow}\label{Ham},\\
  &\lambda_{\mu\nu}\geq 0,\label{condition0}
\end{align}
where $\alpha$ denotes the position of one of the unit plaquettes that
cover the lattice.  $R^{(\mu)}_{\alpha\sigma}$ is an operator whose
expectation value is positive semidefinite
$\braket{R^{(\mu)}_{\alpha\sigma}}\geq 0$. This condition is realized,
if $R^{(\mu)}_{\alpha\sigma}$ is given by a product of an operator and
its Hermitian conjugate. Then the expectation value of the Hamiltonian
is also positive semidefinite $\langle {\cal H}\rangle\geq 0$.

Next, we introduce a trial wave function given by a direct product of
up and down spin sectors,
\begin{equation}
  |\Psi(\mathcal{A},\mathcal{B})\rangle=|\Phi_{\uparrow}(\mathcal{A})\rangle
   \otimes|\Phi_{\downarrow}(\mathcal{B})\rangle,\label{state}
\end{equation}
where $\mathcal{A}$ and $\mathcal{B}$ denote two groups of plaquettes
that cover the lattice satisfying
$\mathcal{A}\cup\mathcal{B}=\{\mbox{all lattice sites}\}$.  We require
that the projection operators have the following conditions,
\begin{equation}
 R^{(\mu)}_{\alpha\uparrow}|\Phi_{\uparrow}(\mathcal{A})\rangle
  =R^{(\mu)}_{\beta\downarrow}|\Phi_{\downarrow}(\mathcal{B})\rangle=0,
  \label{method.5}
\end{equation}
where $\alpha\in\mathcal{A}$ and $\beta\in\mathcal{B}$. Therefore, even
if we have
\begin{equation}
 R^{(\mu)}_{\beta\uparrow}
  |\Phi_{\uparrow}(\mathcal{A})\rangle
  \neq 0,\quad
  R^{(\mu)}_{\alpha\downarrow}
  |\Phi_{\downarrow}(\mathcal{B})\rangle\neq 0,
\end{equation}
the eigenvalue of the Hamiltonian for
$|\Psi(\mathcal{A},\mathcal{B})\rangle$ is always zero.  Then, the lower
bound and the upper bound of the energy are coincide, so that
$|\Psi(\mathcal{A},\mathcal{B})\rangle$ turns out to be one of the exact
ground state of this system.

The above argument can be satisfied in corner sharing lattices with the
bipartite structure. The simplest examples is the 1D
lattice, where the unit plaquette is one bond.  In two dimension,
the Kagom\'e lattice can be covered by two colored triangles
alternatively, as illustrated in Fig.~\ref{fig:lattices}.  These states
can be regarded as the N\'eel ordering on the dual lattice (i.e. the
honeycomb lattice for the Kagom\'e lattice). In three dimension, the
Pyrochlore lattice satisfies these conditions. If the system has a
time-reversal symmetry, its ground state has two-fold degeneracy.

\section{Exact edge states}
\label{sec:exact}

\subsection{1D chain}

We consider the 1D generalized Hubbard model at half-filling and
zero-magnetic field, given by ${\cal H}=\sum_{i\sigma}h_{i,i+1,\sigma}$
with the local bond Hamiltonian,
\begin{align}
 \lefteqn{h_{ij\sigma}=-t\,T_{ij\sigma}
  +\frac{U}{2z}
  (n_{i\sigma}n_{i\bar{\sigma}}+n_{j\sigma}n_{j\bar{\sigma}})
  }\nonumber\\
 &
  +V_{\parallel}n_{i\sigma}n_{j\sigma}+V_{\perp}n_{i\sigma}n_{j\bar{\sigma}}
  \nonumber\\
 &
  +XT_{ij\sigma}(n_{i\bar{\sigma}}+n_{j\bar{\sigma}})
  +\frac{W}{2}\sum_{\sigma'}T_{ij\sigma}T_{ij{\sigma}'},
  \label{local_bond_Ham}
\end{align}
where $\bar{\sigma}$ is the opposite spin of $\sigma$, $z=1$ for the
present 1D case, and periodic boundary conditions are assumed.  We have
defined the hopping and the density operators as
\begin{align}
 T_{ij\sigma}
 \equiv& c_{i\sigma}^{\dag}c_{j\sigma}^{}+\mbox{H.c.},\\
 n_{i\sigma}
 \equiv& c_{i\sigma}^{\dag}c_{i\sigma}^{}.
\end{align}

The exact ground state of the 1D chain has already been discussed in
Refs.~\onlinecite{Itoh-N-M,Nakamura-O-I}.  Here, we discuss parent
Hamiltonians with exact edge states. In the previous section and the
preceding works, we have considered only bulk systems.
In the BN state for the 1D chain,
the local bond Hamiltonian is given by the following form,
\begin{equation}
 h_{ij}-\varepsilon_0(n_i,n_j)=
  \sum_{\mu,\nu}\lambda_{\mu\nu}
  R^{(\mu)}_{ij\uparrow}R^{(\nu)}_{ij\downarrow},\quad
  \lambda_{\mu\nu}\geq 0,
  \label{edge.01}
\end{equation}
where
\begin{equation}
 \varepsilon_0(n_i,n_j)=
  \left(\frac{U}{2}-\frac{W}{2}+t\right)(n_i+n_j-2)+\frac{U}{2}.
  \label{edge.02}
\end{equation}
The right hand side of Eq.~(\ref{edge.01}) gives zero for the BN state
even if edge electrons exist.  The first term of the right hand side of
Eq.~(\ref{edge.02}) vanishes for the half-filling, and the ground state
energy per bond is $U/2$. However, this situation should be modified for
open boundary systems where the sum of the local Hamiltonian becomes
\begin{align}
\lefteqn{\sum_{\braket{i,j}}\left[h_{ij}-\varepsilon_0(n_i,n_j)\right]
  =\mathcal{H}_{\rm bulk}-\frac{U}{2}L}\nonumber\\
 &-\left(\frac{U}{2}-\frac{W}{2}+t\right)
  \left[2(N-L)-n_1-n_{L+1}\right],
  \label{edge.03}
\end{align}
where $L$ is the number of bonds, $N$ is the number of electrons.  $n_1$
and $n_{L+1}$ are the number operators for edge sites.  $\mathcal{H}_{\rm
bulk}$ is the bulk Hamiltonian where the on-site Coulomb
interactions at edges are reduced to the half $U\to U/2$. Then the Hamiltonian
with the exact BN and edge states should be
\begin{equation}
 \mathcal{H}_{\rm edge}=\mathcal{H}_{\rm bulk}
  +\left(\frac{U}{2}-\frac{W}{2}+t\right)(n_1+n_{L+1}),
  \label{edge.04}
\end{equation}
and its ground-state energy is given by
\begin{equation}
 E_0=\frac{U}{2}L+(U-W+2t)n_{\rm edge}
  \label{edge.05}
\end{equation}
where $n_{\rm edge}$ is the number of the localized electrons at the
edges.

\subsection{PN state in Kagom\'e lattice}

We consider the generalized Hubbard model on the Kagom\'e lattice at
$1/3$ and $2/3$-filling with zero-magnetic field. In order to obtain an
exact ground state, we need to include three site terms ($X'$, $W'$
terms).  The Hamiltonian is given by ${\cal H}=\sum_{\langle
ijk\rangle\sigma}h_{ijk\sigma}$, where the summation $\langle
ijk\rangle$ is taken in each unit trimer as shown in
Fig.~\ref{fig:lattices},
\begin{align}
 \lefteqn{h_{ijk\sigma}
 =h_{ij\sigma}+h_{jk\sigma}+h_{ki\sigma}}
  \nonumber\\
 &
  +W'(T_{ij\sigma}T_{jk\bar{\sigma}}+T_{jk\sigma}T_{ki\bar{\sigma}}
  +T_{ki\sigma}T_{ij\bar{\sigma}})\nonumber\\
 &
 +X'(T_{ij\sigma}n_{k\bar{\sigma}}+T_{jk\sigma}n_{i\bar{\sigma}}
 +T_{ki\sigma}n_{j\bar{\sigma}}),
 \label{local_trim_Ham}
\end{align}
where $h_{ij\sigma}$ is the local bond Hamiltonian
(\ref{local_bond_Ham}) with $z=2$. $\bar{\sigma}$ denotes the opposite
spin of $\sigma$.

The exact ground states of the Kagom\'e lattice have already been
discussed in Refs.~\onlinecite{Nakamura-I,Nakamura-N}. Here, we discuss
parent Hamiltonians with exact edge states.  In the case of the PN state
in Kagom\'e lattice at $1/3$ filling, the ground state energy per
plaquette is
\begin{equation}
 \tilde{\varepsilon}_0(N_{ijk})=
  \left(\frac{U}{2}-3W+2t\right)N_{ijk}
  -\left(\frac{U}{2}-4W+4t\right).
  \label{edge.10}
\end{equation}
For bulk systems, it follows from the relation between the number of
plaquettes and the number of sites, $N_{\rm plaq}=(2/3)N_{\rm site}$
that the ground state energy per site becomes
\begin{equation}
 \varepsilon_0=\frac{1}{3}(U-4W).
\end{equation}
For edged systems, the Hamiltonian with the exact PN states should be
\begin{equation}
 \mathcal{H}_{\rm edge}=\mathcal{H}_{\rm bulk}
  +\left(\frac{U}{2}-3W+2t\right)\sum_{i\in{\rm edge}}n_i,
\end{equation}
and its ground-state energy is given by
\begin{equation}
 E_0=\varepsilon_0N_{\rm site}+(U-6W+4t)n_{\rm edge}
  \label{edge.13}
\end{equation}
where $n_{\rm edge}$ is the number of the localized electrons at the
edge.


At $2/3$ filling, the ground state energy per plaquette is
\begin{equation}
 \tilde{\varepsilon}_0(N_{ijk})=
  \left(U-W-t\right)N_{ijk}
  +\left(-2U+4W+4t\right).
  \label{edge.20}
\end{equation}
For bulk systems, it follows from the relation between the number of
plaquettes and the number of sites, $N_{\rm plaq}=(2/3)N_{\rm site}$
that the ground state energy per site becomes
\begin{equation}
 \varepsilon_0=\frac{4}{3}U.
\end{equation}
For edged systems, the Hamiltonian with the exact PN states should be
\begin{equation}
 \mathcal{H}_{\rm edge}=\mathcal{H}_{\rm bulk}
  +\left(U-W-t\right)\sum_{i\in{\rm edge}}n_i,
\end{equation}
and its ground-state energy is given by
\begin{equation}
 E_0=\varepsilon_0N_{\rm site}+2(U-W-t)n_{\rm edge}.
  \label{edge.23}
\end{equation}

\section{Checkerboard lattice}
\label{sec:checker}

In this section, we consider the PN state in the following
generalized Hubbard model on the checkerboard lattice,
\begin{widetext}
\begin{align}
h_{ijkl}-\varepsilon_{0}=&
 -t\sum_{\sigma}T_{ijk\sigma}-\tilde{t}\sum_{\sigma}\tilde{T}_{ijkl\sigma}
 +\frac{U}{2}\sum_{\mu}n_{\mu\uparrow}n_{\mu\downarrow}
 \nonumber \\
&
 +V_{\parallel}\sum_{(\mu,\nu)}\sum_{\sigma} n_{\mu\sigma}n_{\nu\sigma}
 +\tilde{V}_{\parallel}\sum_{\sigma}\left(n_{i\sigma}n_{k\sigma}
 +n_{j\sigma}n_{l\sigma}\right)
 +V_{\perp}\sum_{(\mu,\nu)}\sum_{\sigma}
 n_{\mu,\sigma}n_{\nu,\bar{\sigma}}
 +\tilde{V}_{\perp}\sum_{\sigma}
 \left(n_{i,\sigma}n_{k,\bar{\sigma}}+ n_{j,\sigma}n_{l,\bar{\sigma}}\right)
 \nonumber \\
 &
 +\frac{W}{2}\sum_{(\mu,\nu)}\sum_{\sigma,\sigma'}
 T_{\mu\nu\sigma}T_{\mu\nu\sigma'}
 +\frac{\tilde{W}}{2}\sum_{\sigma,\sigma'}
 \left(T_{ik\sigma}T_{ik\sigma'} + T_{jl\sigma}T_{jl\sigma'}\right)
 \nonumber \\
&
 +X\sum_{(\mu,\nu)}\sum_{\sigma}T_{\mu\nu\sigma}
 \left(n_{\mu\bar{\sigma}} + n_{\nu\bar{\sigma}}\right)
 +X'\sum_{(\mu,\nu,\lambda,\rho)}\sum_{\sigma}T_{\mu\nu\sigma}
 \left(n_{\lambda\bar{\sigma}}+n_{\rho\bar{\sigma}}\right)
 \nonumber \\
 &
 +P\sum_{\sigma}\left(T_{ij\sigma}T_{kl\bar{\sigma}}
 +T_{jk\sigma}T_{li\bar{\sigma}}\right)
 +P'\sum_{\sigma}T_{ik\sigma}T_{jl\bar{\sigma}} \nonumber
\end{align}
where $i,j,k,l$ are taken as indicated in Fig.~\ref{fig:lattices}.
Hereafter, we introduce the constraint $W=W'=V_{\parallel}=P$,
$\tilde{W}=\tilde{V}_{\parallel}=P'$, and $X=X'$.  The plaquette
operators are introduced as
\end{widetext}
\begin{subequations} 
\begin{align}
 A^{\dagger}_{ijkl\sigma}&=\frac{1}{2}(c^{\dagger}_{i\sigma}
 + c^{\dagger}_{j\sigma} + c^{\dagger}_{k\sigma}
 + c^{\dagger}_{l\sigma}),\\
 B^{\dagger}_{ijkl\sigma}&=\frac{1}{2}(c^{\dagger}_{i\sigma}
 + \i c^{\dagger}_{j\sigma} - c^{\dagger}_{k\sigma}
 - \i c^{\dagger}_{l\sigma}),\\
 C^{\dagger}_{ijkl\sigma}&=\frac{1}{2}(c^{\dagger}_{i\sigma}
 - \i c^{\dagger}_{j\sigma} - c^{\dagger}_{k\sigma}
 + \i c^{\dagger}_{l\sigma}),\\
 D^{\dagger}_{ijkl\sigma}&=\frac{1}{2}(c^{\dagger}_{i\sigma}
 - c^{\dagger}_{j\sigma} + c^{\dagger}_{k\sigma}
 - c^{\dagger}_{l\sigma}).
\end{align}
\end{subequations}
The plaquette operators in the same plaquette satisfy the
anticommutation relations:
\begin{align}
 &\{A_{ijkl\sigma}^{\mathstrut},A^{\dagger}_{ijkl\sigma'}\}
 =\{B_{ijkl\sigma}^{\mathstrut},B^{\dagger}_{ijkl\sigma'}\}
 =\{C_{ijkl\sigma}^{\mathstrut},C^{\dagger}_{ijkl\sigma'}\} \nonumber \\
 &=\{D_{ijkl\sigma}^{\mathstrut},D^{\dagger}_{ijkl\sigma'}\}
 =\delta_{\sigma \sigma'},
\end{align}
and other anticommutators are zero.  The density operators of the
plaquette operators is defined by
\begin{subequations} 
\begin{align}
 n_{A\sigma} &= A^{\dagger}_{ijkl\sigma}A_{ijkl\sigma}
 =\frac{1}{4}(N_{ijkl\sigma} + T_{ijkl\sigma} + T'_{ijkl\sigma}),\\
 n_{B\sigma} &= B^{\dagger}_{ijkl\sigma}B_{ijkl\sigma}
 =\frac{1}{4}(N_{ijkl\sigma} - T'_{ijkl\sigma} - J_{ijkl\sigma}),\\
 n_{C\sigma} &= C^{\dagger}_{ijkl\sigma}C_{ijkl\sigma}
 =\frac{1}{4}(N_{ijkl\sigma} - T'_{ijkl\sigma} + J_{ijkl\sigma}),\\
 n_{D\sigma} &= D^{\dagger}_{ijkl\sigma}D_{ijkl\sigma}
 =\frac{1}{4}(N_{ijkl\sigma} - T_{ijkl\sigma} + T'_{ijkl\sigma}),
\end{align}
\end{subequations}
where density, hopping and current operators are defined as follows
\begin{subequations} 
\begin{align}
 N_{ijkl\sigma}&=n_{i\sigma}+n_{j\sigma}+n_{k\sigma}+n_{l\sigma},\\
 T_{ijkl\sigma}&=T_{ij\sigma}+T_{jk\sigma}+T_{kl\sigma}+T_{li\sigma},\\
 \tilde{T}_{ijkl\sigma}&=T_{ik\sigma}+T_{jl\sigma},\\
 J_{ijkl\sigma}&=J_{ij\sigma}+J_{jk\sigma}+J_{kl\sigma}+J_{li\sigma}.
\end{align}
\end{subequations}

\begin{figure*}[t]
\begin{center}
 \includegraphics[width=170mm]{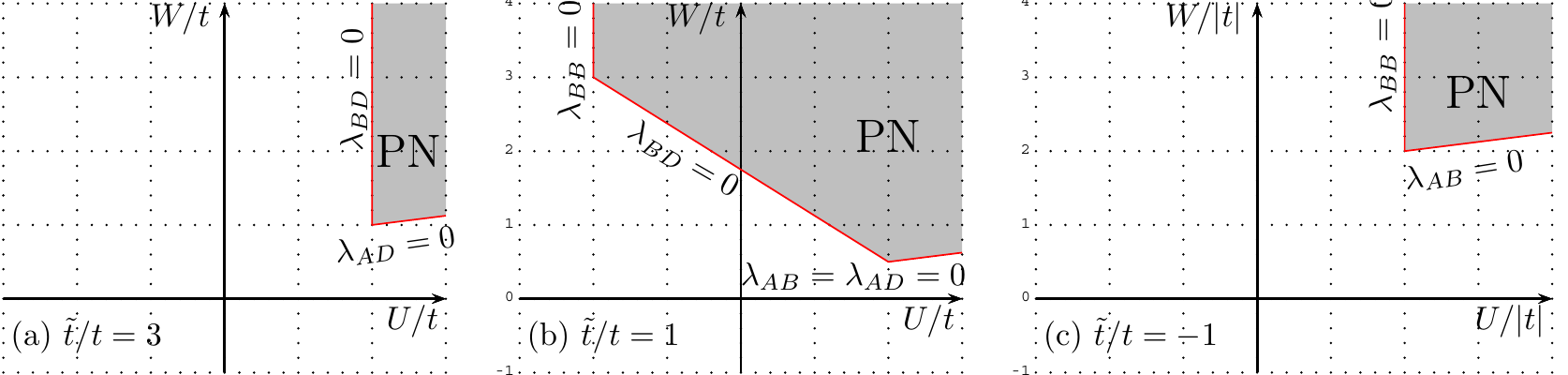}
 \end{center}
 \caption{Phase diagrams of the generalized Hubbard model on the
 checkerboard lattice at 1/4 filling for several values of
 $\tilde{t}/t$. The shaded region is the PN state.}
 \label{fig:checkerboard1}
\end{figure*}

\subsection{Plaquette-N\'eel state at 1/4-filling}

First, we consider the PN state at 1/4-filling.  This
state is given by
\begin{equation}
 \Ket{\Psi_{\sigma}}=\prod_{\braket{ijkl}\in \Box}A^{\dagger}_{ijkl\sigma}
  \prod_{\braket{i'j'k'l'} \in \Box'}A^{\dagger}_{i'j'k'l'\bar{\sigma}}\Ket{0},
\end{equation}
where the sum $\braket{ijkl}$ ($\braket{i'j'k'l'}$) is taken for all
blue (red) plaquettes of the checkerboard lattice in
Fig.~\ref{fig:lattices}.  The parent Hamiltonian for this state is
constructed as
\begin{align}
 &h_{ijkl} - \tilde{\varepsilon}_{0}
 =\lambda_{AA}(1-n_{A\uparrow})(1-n_{A\downarrow})
 + \lambda_{BB}n_{B\uparrow}n_{B\downarrow} \nonumber \\
 &+ \lambda_{CC}n_{C\uparrow}n_{C\downarrow}
 + \lambda_{DD}n_{D\uparrow}n_{D\downarrow}  \nonumber \\
 &~~+ \lambda_{AB}\left[(1-n_{A\uparrow})n_{B\downarrow}
 + n_{B\uparrow}(1-n_{A\downarrow})\right] \nonumber \\
 &+ \lambda_{AC}\left[(1-n_{A\uparrow})n_{C\downarrow}
 + n_{C\uparrow}(1-n_{A\downarrow})\right]  \nonumber \\
 &~~~~+ \lambda_{AD}\left[(1-n_{A\uparrow})n_{D\downarrow}
 + n_{D\uparrow}(1-n_{A\downarrow})\right] \nonumber \\
 &+ \lambda_{BC}\left[n_{B\uparrow}n_{C\downarrow}
 + n_{C\uparrow}n_{B\downarrow}\right] \nonumber \\
 &~~~~~~+ \lambda_{BD}\left[n_{B\uparrow}n_{D\downarrow}
 + n_{D\uparrow}n_{B\downarrow}\right] \nonumber \\
 & + \lambda_{CD}\left[n_{C\uparrow}n_{D\downarrow}
 + n_{D\uparrow}n_{C\downarrow}\right].
\end{align}
Here, we set the parameters assuming time-reversal symmetry as
\begin{equation}
 \lambda_{BB}=\lambda_{CC}=\lambda_{BC},\quad
 \lambda_{AB}=\lambda_{AC},\quad\lambda_{BD}=\lambda_{CD}.
 \label{setpara}
\end{equation}
Then the relations
between $\lambda$ and the parameters of the Hamiltonian are identified
as
\begin{align}
 \lambda_{AA}&=4t+\tilde{t}+\frac{1}{2}U-4W,\nonumber\\
 \lambda_{BB}&=\tilde{t}+\frac{1}{2}U,\nonumber\\
 \lambda_{DD}&=-4t+\tilde{t}+\frac{1}{2}U+12W,\nonumber\\
 \lambda_{AB}&=-2t+2\tilde{t}-\frac{1}{2}U+4W,\\
 \lambda_{AD}&=-\tilde{t}-\frac{1}{2}U + 4W,\nonumber\\
 \lambda_{BD}&=-2t-\tilde{t}+\frac{1}{2}U+4W,\nonumber
\end{align}
with the relations,
\begin{equation}
 \tilde{W}=\tilde{t},\qquad
 X=t-2W.
\end{equation}
The energy per plaquette is given by
\begin{align}
\tilde{\varepsilon}_{0}
 =&-4t-\tilde{t}-\frac{1}{2}U+4W\nonumber\\
 &+\frac{1}{2}(4t+\tilde{t}+U-6W)\sum_{\sigma}N_{ijkl\sigma}.
\end{align}
For bulk systems, it follows from the relation between the number of
plaquettes and the number of sites, $N_{\rm plaq}=(1/2)N_{\rm site}$
that the ground state energy per site becomes
\begin{equation}
 \varepsilon_{0}=\frac{1}{4}U-W.
\end{equation}
For edged systems, the Hamiltonian with the exact PN states should be
\begin{equation}
 \mathcal{H}_{\rm edge}=\mathcal{H}_{\rm bulk}
  +\frac{1}{2}(4t+\tilde{t}+U-6W)\sum_{i\in{\rm edge}}n_i,
\end{equation}
and its ground-state energy is given by
\begin{equation}
 E_0=\varepsilon_0N_{\rm site}+(4t+\tilde{t}+U-6W)n_{\rm edge}
\end{equation}
where $n_{\rm edge}$ is the number of the localized electrons at the
edge.

The conditions of this state is given as follows,
\begin{align}
 \frac{W}{t} &\leq
 1+\frac{1}{4}\frac{\tilde{t}}{t}+\frac{1}{8}\frac{U}{t},
 \nonumber\\
 U &\geq -2\tilde{t},\nonumber\\
 \frac{W}{t} &\geq \frac{1}{3}-\frac{1}{12}\frac{\tilde{t}}{t}
 -\frac{1}{24}\frac{U}{t},\nonumber\\
 \frac{W}{t} &\geq \frac{1}{2}-\frac{1}{2}\frac{\tilde{t}}{t}
 +\frac{1}{8}\frac{U}{t},\\
 \frac{W}{t} &\geq \frac{1}{4}\frac{\tilde{t}}{t}
 +\frac{1}{8}\frac{U}{t},\nonumber\\
 \frac{W}{t} &\geq \frac{1}{2}+\frac{1}{4}\frac{\tilde{t}}{t}
 -\frac{1}{8}\frac{U}{t}.\nonumber
\end{align}
Then we obtain the phase diagrams of this state for $t>0$ and $t<0$
regions as shown in Fig.~\ref{fig:checkerboard1}.

\begin{figure*}[t]
 \begin{center}
 \includegraphics[width=170mm]{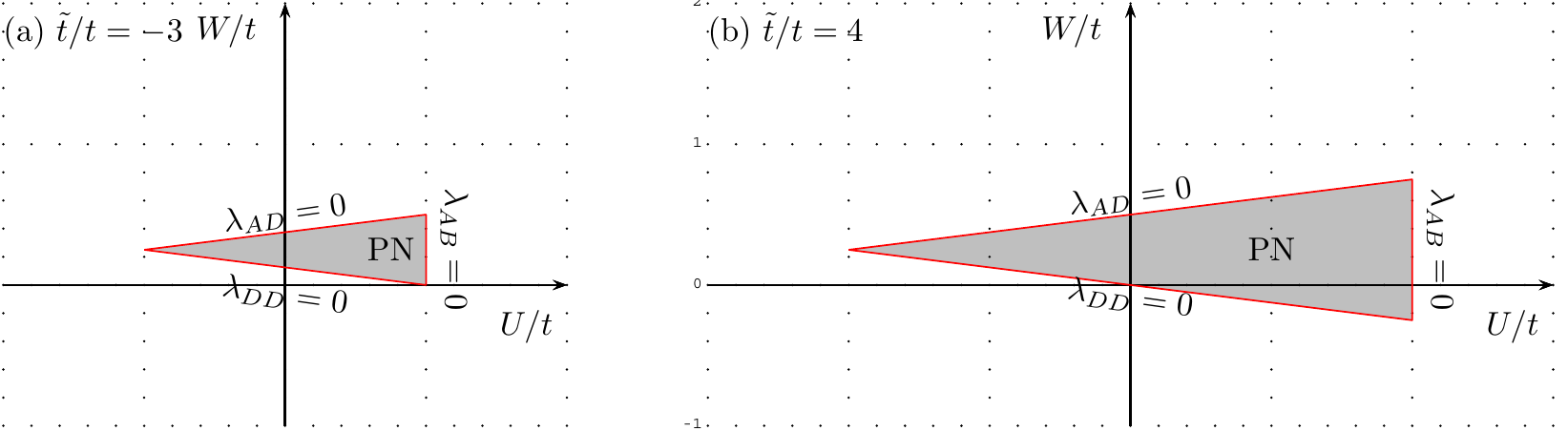}
 \end{center}
\caption{Phase diagrams of the generalized Hubbard model on the
 checkerboard lattice at 1/2 filling for (a) $\ket{CB}$ state and (b)
 $\ket{DA}$ state for $t>0$. The shaded regions are the PN
 states.}  \label{fig:checkerboard2}
\end{figure*}

\subsection{Plaquette-N\'eel state at 1/2-filling}

\subsubsection{$\Ket{CB}$ state}

Next we consider the PN state at 1/2-filling, which is
given as two electrons in each plaquette,
\begin{equation}
 \Ket{\Psi_{\sigma}}=\prod_{\braket{ijkl} \in \Box}
  C^{\dagger}_{ijkl\sigma}B^{\dagger}_{ijkl\sigma}
  \prod_{\braket{i'j'k'l'} \in \Box'}
  C^{\dagger}_{i'j'k'l'\bar{\sigma}}B^{\dagger}_{i'j'k'l'\bar{\sigma}}
  \Ket{0}.
\end{equation}
Since $B^{\dagger}_{ijkl\sigma}$ and $C^{\dagger}_{ijkl\sigma}$ creates
flux states with opposite directions, the state $\Ket{\Psi_{\sigma}}$
has no local current.  The Hamiltonian for this state is constructed as
\begin{align}
 &h_{ijkl} - \tilde{\varepsilon}_{0}=\lambda_{AA}n_{A\uparrow}n_{A\downarrow}
 + \lambda_{BB}(1-n_{B\uparrow})(1-n_{B\downarrow}) \nonumber \\
 &+ \lambda_{CC}(1-n_{C\uparrow})(1-n_{C\downarrow})
 + \lambda_{DD}n_{D\uparrow}n_{D\downarrow} \nonumber \\
 &+ \lambda_{AB}\left[n_{A\uparrow}(1-n_{B\downarrow})
 + (1-n_{B\uparrow})n_{A\downarrow}\right] \nonumber \\
 &+ \lambda_{AC}\left[n_{A\uparrow}(1-n_{C\downarrow})
 + (1-n_{C\uparrow})n_{A\downarrow}\right] \nonumber \\
 &+ \lambda_{AD}\left[n_{A\uparrow}n_{D\downarrow}
 + n_{D\uparrow}n_{A\downarrow}\right] \nonumber \\
 &+ \lambda_{BC}\left[(1-n_{B\uparrow})(1-n_{C\downarrow})
 + (1-n_{C\uparrow})(1-n_{B\downarrow})\right] \nonumber \\
 &+ \lambda_{BD}\left[(1-n_{B\uparrow})n_{D\downarrow}
 + n_{D\uparrow}(1-n_{B\downarrow})\right] \nonumber \\
 &+ \lambda_{CD}\left[(1-n_{C\uparrow})n_{D\downarrow}
 + n_{D\uparrow}(1-n_{C\downarrow})\right].
\end{align}
We set the parameters as Eq.~(\ref{setpara}) assuming time-reversal
symmetry.  Then the relations between $\lambda$ and parameters of the
Hamiltonian are identified as
\begin{align}
 \lambda_{AA}&=2t-\frac{1}{2}\tilde{t}+\frac{1}{2}U+4W,\nonumber\\
 \lambda_{BB}&=-\frac{1}{2}\tilde{t}+\frac{1}{2}U,\nonumber\\
 \lambda_{DD}&=-2t-\frac{1}{2}\tilde{t}+\frac{1}{2}U+4W,\nonumber\\
 \lambda_{AB}&=-t-\frac{1}{2}\tilde{t}-\frac{1}{2}U,\\
 \lambda_{AD}&=-\frac{1}{2}\tilde{t}+\frac{1}{2}U-4W,\nonumber\\
 \lambda_{BD}&=t-\frac{1}{2}t-\frac{1}{2}U,\nonumber
\end{align}
with the relations,
\begin{equation}
 X=\frac{1}{2}t,\qquad
 \tilde{W}=-\frac{1}{2}\tilde{t}.
\end{equation}
The energy per plaquette is
\begin{equation}
 \tilde{\varepsilon}_{0}
=2\tilde{t}-2U
 +\frac{1}{4}(-\tilde{t}+4U+4W)\sum_{\sigma}N_{ijkl\sigma}
\end{equation}
For bulk systems, it follows from the relation between the number of
plaquettes and the number of sites, $N_{\rm plaq}=(1/2)N_{\rm site}$
that the ground state energy per site becomes
\begin{equation}
 \varepsilon_{0}=\frac{1}{2}(\tilde{t}+2U+4W).
\end{equation}
For edged systems, the Hamiltonian with the exact PN states should be
\begin{equation}
 \mathcal{H}_{\rm edge}=\mathcal{H}_{\rm bulk}
  +\frac{1}{4}(-\tilde{t}+4U+4W)\sum_{i\in{\rm edge}}n_i
\end{equation}
and its ground-state energy is given by
\begin{equation}
 E_0=\varepsilon_0N_{\rm site}
  +\frac{1}{2}(-\tilde{t}+4U+4W)n_{\rm edge}
\end{equation}
where $n_{\rm edge}$ is the number of the localized electrons at the
 edge.

Thus the condition of the exact ground state is given by
\begin{align}
 W &\geq -\frac{1}{2}t+\frac{1}{8}\tilde{t}-\frac{1}{8}U,\nonumber\\
 U &\geq \tilde{t},\nonumber\\
 W &\geq \frac{1}{2}t+\frac{1}{8}\tilde{t}-\frac{1}{8}U,\nonumber\\
 U &\leq -2t-\tilde{t},\\
 W &\leq -\frac{1}{8}\tilde{t}+\frac{1}{8}U,\nonumber\\
 U  &\leq 2t-\tilde{t}.\nonumber
\end{align}
Then we obtain the phase diagram of the PN state for $\tilde{t}/t<2$
($t>0$) and $\tilde{t}/|t|<-2$ ($t<0$) regions as shown in
Fig.~\ref{fig:checkerboard2}(a).

\subsubsection{$\Ket{DA}$ state}

Similarly, we consider the PN state given by $\Ket{DA}$
state at 1/2-filling,
\begin{equation}
 \Ket{\Psi_{\sigma}}=\prod_{\braket{ijkl} \in \Box}
  D^{\dagger}_{ijkl\sigma}A^{\dagger}_{ijkl\sigma}
  \prod_{\braket{i'j'k'l'} \in \Box'}
  D^{\dagger}_{i'j'k'l'\bar{\sigma}}A^{\dagger}_{i'j'k'l'\bar{\sigma}}\Ket{0},
\end{equation}
the relations between $\lambda$ and parameters of the Hamiltonian are
identified as
\begin{align}
\lambda_{AA}=&2t+\frac{1}{2}\tilde{t}+\frac{1}{2}U+4W, \nonumber \\
\lambda_{BB}=&\frac{1}{2}\tilde{t}+\frac{1}{2}U, \nonumber \\
\lambda_{DD}=&-2t+\frac{1}{2}\tilde{t}+\frac{1}{2}U+4W, \nonumber \\
\lambda_{AB}=&-t+\frac{1}{2}\tilde{t}-\frac{1}{2}U,\\
\lambda_{AD}=&\frac{1}{2}\tilde{t}+\frac{1}{2}U-4W, \nonumber \\
\lambda_{BD}=&t+\frac{1}{2}\tilde{t}-\frac{1}{2}U, \nonumber
\end{align}
with
\begin{equation}
\tilde{W}=\frac{1}{2}\tilde{t},\qquad
X=\frac{1}{2}t.
\end{equation}
The energy per plaquette is given by
\begin{equation}
\tilde{\varepsilon}_{0}
 =-2\tilde{t}-2U+\frac{1}{4}(\tilde{t}+4U+4W)
 \sum_{\sigma}N_{ijkl\sigma}
\end{equation}
For bulk systems, it follows from the relation between the number of
plaquettes and the number of sites, $N_{\rm plaq}=(1/2)N_{\rm site}$
that the ground state energy per site becomes
\begin{equation}
 \varepsilon_{0}=\frac{1}{2}(\tilde{t}+2U+4W).
\end{equation}
For edged systems, the Hamiltonian with the exact PN states should be
\begin{equation}
 \mathcal{H}_{\rm edge}=\mathcal{H}_{\rm bulk}
  +\frac{1}{4}(\tilde{t}+4U+4W)\sum_{i\in{\rm edge}}n_i
\end{equation}
and its ground-state energy is given by
\begin{equation}
 E_0=\varepsilon_0N_{\rm site}
  +\frac{1}{2}(\tilde{t}+4U+4W)n_{\rm edge}
\end{equation}
where $n_{\rm edge}$ is the number of the localized electrons at the
 edge.

Thus conditions of this state is given as follows,
\begin{align}
\frac{W}{t} \geq&
 -\frac{1}{2}-\frac{1}{8}\frac{\tilde{t}}{t}-\frac{1}{8}\frac{U}{t},
 \nonumber \\
\frac{U}{t} \geq&
 -\frac{\tilde{t}}{t}, \nonumber \\
\frac{W}{t} \geq&
 \frac{1}{2}-\frac{1}{8}\frac{\tilde{t}}{t}-\frac{1}{8}\frac{U}{t}, \\
\frac{U}{t} \leq&
 -2+\frac{\tilde{t}}{t}, \nonumber \\
\frac{W}{t} \leq&
\frac{1}{8}\frac{\tilde{t}}{t}+\frac{1}{8}\frac{U}{t}, \nonumber \\
\frac{U}{t} \leq&
 2+\frac{\tilde{t}}{t}. \nonumber
\end{align}
Then we obtain the phase diagram of this state for $\tilde{t}/t<2$
($t>0$) and $\tilde{t}/|t|<-2$ ($t<0$) regions as shown in
Fig.~\ref{fig:checkerboard2}(b).

\subsection{Plaquette-N\'eel state at 3/4-filling}

\subsubsection{$\Ket{CBA}$ state}

We consider the PN state at 3/4-filling, which is given as
three electrons in each plaquette,
\begin{align}
\Ket{\Psi_{\sigma}}=&\prod_{\braket{ijkl} \in \Box}
 C^{\dagger}_{ijkl\sigma}
 B^{\dagger}_{ijkl\sigma}
 A^{\dagger}_{ijkl\sigma} \nonumber \\
 &\times \prod_{\braket{i'j'k'l'} \in \Box'}
 C^{\dagger}_{i'j'k'l'\bar{\sigma}}
 B^{\dagger}_{i'j'k'l'\bar{\sigma}}
 A^{\dagger}_{i'j'k'l'\bar{\sigma}}\Ket{0},
\end{align}
where the sum $\braket{ijkl}$ ($\braket{i'j'k'l'}$) is taken for all
blue (red) plaquettes of the checkerboard lattice in
Fig.~\ref{fig:lattices}.  The parent Hamiltonian for this state is
constructed as
\begin{align}
&h_{ijkl} - \tilde{\varepsilon}_{0}
 =\lambda_{AA}(1-n_{A\uparrow})(1-n_{A\downarrow}) \nonumber \\
 &+ \lambda_{BB}(1-n_{B\uparrow})(1-n_{B\downarrow}) \nonumber \\
 &+ \lambda_{CC}(1-n_{C\uparrow})(1-n_{C\downarrow})
 + \lambda_{DD}n_{D\uparrow}n_{D\downarrow}  \nonumber \\
 &+ \lambda_{AB}\left[(1-n_{A\uparrow})(1-n_{B\downarrow})
 + (1-n_{B\uparrow})(1-n_{A\downarrow})\right]  \nonumber \\
 &+ \lambda_{AC}\left[(1-n_{A\uparrow})(1-n_{C\downarrow})
 + (1-n_{C\uparrow})(1-n_{A\downarrow})\right]  \nonumber \\
 &+ \lambda_{AD}\left[(1-n_{A\uparrow})n_{D\downarrow}
 + n_{D\uparrow}(1-n_{A\downarrow})\right]  \nonumber \\
 &+ \lambda_{BC} \left[(1-n_{B\uparrow})(1-n_{C\downarrow})
 + (1-n_{C\uparrow})(1-n_{B\downarrow})\right] \nonumber \\
 &+ \lambda_{BD}\left[(1-n_{B\uparrow})n_{D\downarrow}
 + n_{D\uparrow}(1-n_{B\downarrow})\right] \nonumber \\
 &+ \lambda_{CD}\left[(1-n_{C\uparrow})n_{D\downarrow}
 + n_{D\uparrow}(1-n_{C\downarrow})\right].
\end{align}
We set the parameters as Eq.~(\ref{setpara}) assuming time-reversal
symmetry.  Then relations between $\lambda$ and the parameters of the
Hamiltonian are identified as
\begin{align}
\lambda_{AA}=&\frac{4}{3}t-\tilde{t}+\frac{1}{2}U+\frac{4}{3}W,\nonumber\\
\lambda_{BB}=&-\tilde{t}+\frac{1}{2}U, \nonumber \\
\lambda_{DD}=&-\frac{4}{3}t-\tilde{t}+\frac{1}{2}U+\frac{20}{3}W,\nonumber\\
\lambda_{AB}=&\frac{2}{3}t+\tilde{t}+\frac{1}{2}U-\frac{4}{3}W,\nonumber \\
\lambda_{AD}=&\tilde{t}-\frac{1}{2}U+4W,\nonumber \\
\lambda_{BD}=&\frac{2}{3}t-\tilde{t}-\frac{1}{2}U-\frac{4}{3}W,\nonumber
\end{align}
with the relations
\begin{equation}
\tilde{W}=-\tilde{t},\qquad
X=\frac{1}{3}t-\frac{2}{3}W.
\end{equation}
The energy per plaquette is
\begin{align}
\tilde{\varepsilon}_{0}
=&-4t+\tilde{t}-\frac{9}{2}U+4W\nonumber\\
 &+\frac{1}{6}(4t-3\tilde{t}+9U-2W)
 \sum_{\sigma}N_{ijkl\sigma}.
\end{align}
For bulk systems, it follows from the relation between the number of
plaquettes and the number of sites, $N_{\rm plaq}=(1/2)N_{\rm site}$
that the ground state energy per site becomes
\begin{equation}
 \varepsilon_{0}=-\tilde{t}+\frac{9}{4}U+W.
\end{equation}
For edged systems, the Hamiltonian with the exact PN states should be
\begin{equation}
 \mathcal{H}_{\rm edge}=\mathcal{H}_{\rm bulk}
  +\frac{1}{6}(4t-3\tilde{t}+9U-2W)\sum_{i\in{\rm edge}}n_i
\end{equation}
and its ground-state energy is given by
\begin{equation}
 E_0=\varepsilon_0N_{\rm site}
  +\frac{1}{3}(4t-3\tilde{t}+9U-2W)n_{\rm edge}
\end{equation}
where $n_{\rm edge}$ is the number of the localized electrons at the
 edge.

\begin{figure}[t]
\begin{center}
 \includegraphics[width=80mm]{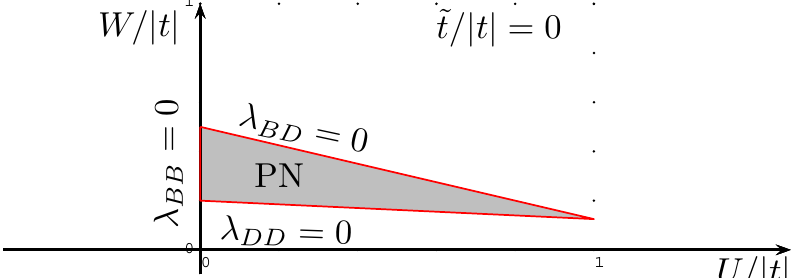}
 \end{center}
\caption{Phase diagrams of the generalized Hubbard model on the
 checkerboard lattice at 3/4 filling. The shaded region is the PN
 state.}  \label{fig:checkerboard3}
\end{figure}

The condition for the exact PN state with $t>0$ is given by $\lambda\geq
0$ as
\begin{align}
 \frac{W}{t} \geq
 & -1+\frac{3}{4}\frac{\tilde{t}}{t}-\frac{3}{8}\frac{U}{t}, \nonumber \\
 \frac{U}{t} \geq& 2\frac{\tilde{t}}{t}, \nonumber \\
 \frac{W}{t} \geq
 & \frac{1}{5}+\frac{3}{20}\frac{\tilde{t}}{t}-\frac{3}{40}\frac{U}{t},
 \nonumber \\
 \frac{W}{t} \leq& \frac{1}{2}+\frac{3}{4}\frac{\tilde{t}}{t}
 +\frac{3}{8}\frac{U}{t}, \nonumber \\
 \frac{W}{t} \geq& -\frac{1}{4}\frac{\tilde{t}}{t}
 +\frac{1}{8}\frac{U}{t}, \nonumber \\
 \frac{W}{t} \leq& \frac{1}{2}-\frac{3}{4}\frac{\tilde{t}}{t}
 -\frac{3}{8}\frac{U}{t}.
\end{align}
There is a finite region for $-1<\tilde{t}/t<1/5$ as shown in
Fig.~\ref{fig:checkerboard3}. On the other hand, there is no parameter
region for $t<0$.

\subsubsection{$\Ket{DCB}$ state}

Similarly, we consider the PN state given by $\Ket{DCB}$
state at 3/4-filling,
\begin{align}
\Ket{\Psi_{\sigma}}=&\prod_{\braket{ijkl} \in \Box}
 D^{\dagger}_{ijkl\sigma}C^{\dagger}_{ijkl\sigma}
 B^{\dagger}_{ijkl\sigma} \nonumber \\
 &\times \prod_{\braket{i'j'k'l'} \in \Box'}
 D^{\dagger}_{i'j'k'l'\bar{\sigma}}
 C^{\dagger}_{i'j'k'l'\bar{\sigma}}
 B^{\dagger}_{i'j'k'l'\bar{\sigma}}\Ket{0}.
\end{align}
Then relations between $\lambda$ and the parameters of the Hamiltonian
are identified as
\begin{align}
 \lambda_{AA}&=\frac{4}{3}t-\tilde{t}+\frac{1}{2}U+\frac{20}{3}W,\nonumber\\
 \lambda_{BB}&=-\tilde{t}+\frac{1}{2}U,\nonumber\\
 \lambda_{DD}&=-\frac{4}{3}t-\tilde{t}+\frac{1}{2}U+\frac{4}{3}W,\nonumber\\
 \lambda_{AB}&=-\frac{2}{3}t-\tilde{t}-\frac{1}{2}U-\frac{4}{3}W,\\
 \lambda_{AD}&=\tilde{t}-\frac{1}{2}U+4W,\nonumber\\
 \lambda_{BD}&=-\frac{2}{3}t+\tilde{t}+\frac{1}{2}U-\frac{4}{3}W,\nonumber
\end{align}
with the following relations
\begin{equation}
 \tilde{W}=-\tilde{t},\qquad X=\frac{1}{3}t-\frac{2}{3}W.
\end{equation}
The energy per plaquette is
\begin{align}
\tilde{\varepsilon}_{0}
=&-4t+\tilde{t}-\frac{9}{2}U+4W\nonumber\\
 &+\frac{1}{6}(4t-3\tilde{t}+9U-2W)
 \sum_{\sigma}N_{ijkl\sigma},
\end{align}
where $t$ is negative, because
$\lambda_{DD}+2\lambda_{AB}+\lambda_{AD}+2\lambda_{BD}=-4t$.

For bulk systems, it follows from the relation between the number of
plaquettes and the number of sites, $N_{\rm plaq}=(1/2)N_{\rm site}$
that the ground state energy per site becomes
\begin{equation}
 \varepsilon_{0}=-\tilde{t}+\frac{9}{4}U+W.
\end{equation}
For edged systems, the Hamiltonian with the exact PN states should be
\begin{equation}
 \mathcal{H}_{\rm edge}=\mathcal{H}_{\rm bulk}
  +\frac{1}{6}(4t-3\tilde{t}+9U-2W)\sum_{i\in{\rm edge}}n_i
\end{equation}
and its ground-state energy is given by
\begin{equation}
 E_0=\varepsilon_0N_{\rm site}
  +\frac{1}{3}(4t-3\tilde{t}+9U-2W)n_{\rm edge}
\end{equation}
where $n_{\rm edge}$ is the number of the localized electrons at the
edge.

 Finally one finds that the PN region is given by the same as that of
the $\ket{CBA}$ state with $t\to -t$.

\section{Pyrochlore lattice}
\label{sec:pyro}

In the case of the pyrochlore lattice, the generalized Hubbard model is
given by
\begin{widetext}
\begin{align}
 h_{ijkl}=&-t\sum_{\sigma}T_{ijk\sigma}
+U\frac{1}{2}\sum_{\mu}n_{\mu\uparrow}n_{\mu\downarrow}
+V_{\perp}\sum_{(\mu,\nu)}\sum_{\sigma}
 n_{\mu,\sigma}n_{\nu,\bar{\sigma}}
+V_{\parallel}\sum_{(\mu,\nu)}\sum_{\sigma} n_{\mu\sigma}n_{\nu\sigma}
\nonumber\\
&+W\frac{1}{2}\sum_{(\mu,\nu)}\sum_{\sigma,\sigma'}
T_{\mu\nu\sigma}T_{\mu\nu\sigma'}
+W'\sum_{(\mu,\nu,\lambda)}\sum_{\sigma}T_{\mu\nu\sigma}
T_{\nu\lambda\bar{\sigma}}
+P\sum_{\sigma}\left(T_{ij\sigma}T_{kl\bar{\sigma}}
 + T_{jk\sigma}T_{li\bar{\sigma}}
 + T_{ik\sigma}T_{jl\bar{\sigma}}\right)\nonumber\\
&+X\sum_{(\mu,\nu)}\sum_{\sigma}T_{\mu\nu\sigma}
 \left(n_{\mu\bar{\sigma}} + n_{\nu\bar{\sigma}}\right)
+X'\sum_{(\mu,\nu,\lambda,\rho)}\sum_{\sigma}T_{\mu\nu\sigma}
\left(n_{\lambda\bar{\sigma}}+n_{\rho\bar{\sigma}}\right)
\end{align}
\end{widetext}
and the plaquette operators are introduced as
\begin{subequations} 
\begin{align}
 A^{\dagger}_{ijkl\sigma}&=\frac{1}{2}(c^{\dagger}_{i\sigma}
 + c^{\dagger}_{j\sigma} + c^{\dagger}_{k\sigma} + c^{\dagger}_{l\sigma}),\\
 B^{\dagger}_{ijkl\sigma}&=\frac{1}{2}(-c^{\dagger}_{i\sigma}
 + c^{\dagger}_{j\sigma} + c^{\dagger}_{k\sigma} - c^{\dagger}_{l\sigma}),\\
 C^{\dagger}_{ijkl\sigma}&=\frac{1}{2}(-c^{\dagger}_{i\sigma}
 + c^{\dagger}_{j\sigma} - c^{\dagger}_{k\sigma} + c^{\dagger}_{l\sigma}),\\
 D^{\dagger}_{ijkl\sigma}&=\frac{1}{2}(-c^{\dagger}_{i\sigma}
 - c^{\dagger}_{j\sigma} + c^{\dagger}_{k\sigma} + c^{\dagger}_{l\sigma}).
\end{align}
\end{subequations}
These definitions are different from those of the checkerboard lattice
reflecting the symmetry of the tetrahedra.  The plaquette operators on
the same tetrahedra satisfy the anticommutation relations:
\begin{align}
 &\{A_{ijkl\sigma},A^{\dagger}_{ijkl\sigma'}\}
 =\{B_{ijkl\sigma},B^{\dagger}_{ijkl\sigma'}\}
 =\{C_{ijkl\sigma},C^{\dagger}_{ijkl\sigma'}\}\nonumber\\
 &=\{D_{ijkl\sigma},D^{\dagger}_{ijkl\sigma'}\}=\delta_{\sigma \sigma'},
\end{align}
and other anticommutators are zero.  The density operators of the
plaquette operators are
\begin{subequations} 
\begin{align}
 n_{A\sigma} &= A^{\dagger}_{ijkl\sigma}A_{ijkl\sigma}
 =\frac{1}{4}(N_{ijkl\sigma} + T_{ijkl\sigma} ),\\
 n_{B\sigma} &= B^{\dagger}_{ijkl\sigma}B_{ijkl\sigma}\nonumber\\
 &=\frac{1}{4}\left\{N_{ijkl\sigma} - T_{ijkl\sigma}
 + 2(T_{li\sigma} + T_{jk\sigma}) \right\},\\
 n_{C\sigma} &= C^{\dagger}_{ijkl\sigma}C_{ijkl\sigma}\nonumber\\
 &=\frac{1}{4}\left\{N_{ijkl\sigma} - T_{ijkl\sigma}
 + 2(T_{ik\sigma} + T_{jl\sigma}) \right\},\\
 n_{D\sigma} &= D^{\dagger}_{ijkl\sigma}D_{ijkl\sigma}\nonumber\\
 &=\frac{1}{4}\left\{N_{ijkl\sigma} - T_{ijkl\sigma}
 + 2(T_{ij\sigma} + T_{kl\sigma}) \right\},
\end{align}
\end{subequations}
where the density, the hopping and the current operators are defined as
follows
\begin{align}
 N_{ijkl\sigma}&=n_{i\sigma}+n_{j\sigma}+n_{k\sigma}+n_{l\sigma},\\
 T_{ijkl\sigma}&=T_{ij\sigma}+T_{jk\sigma}+T_{kl\sigma}.
 +T_{li\sigma}+T_{ik\sigma}+T_{jl\sigma}.
\end{align}

\begin{figure*}
\begin{center}
 \includegraphics[width=170mm]{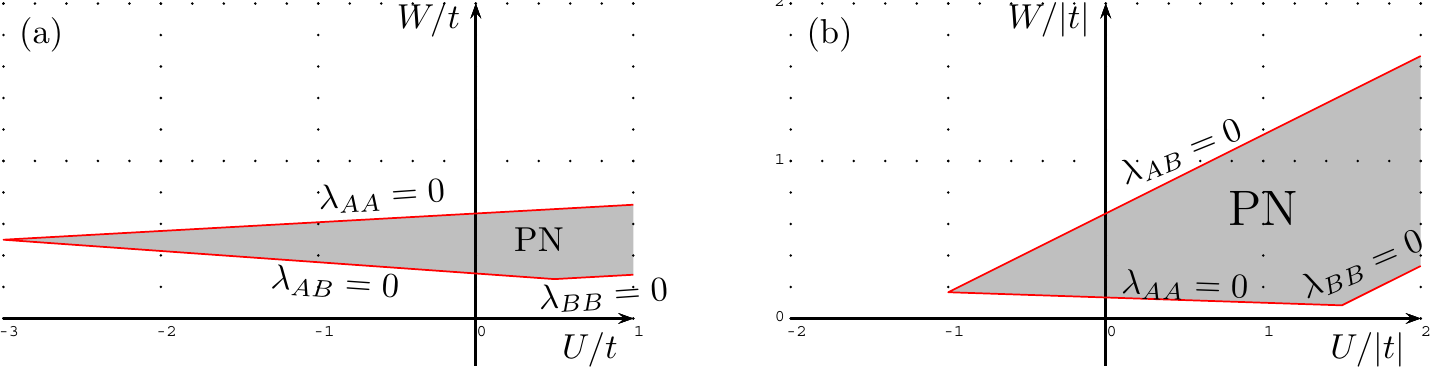}
\end{center}
\caption{Phase diagrams of the generalized Hubbard model on the
 pyrochlore lattice at (a) 1/4 filling and (b) 3/4 filling. The shaded
 regions are the PN states.}
 \label{fig:pyrochlore}
\end{figure*}

\subsection{Plaquette-N\'eel state at 1/4-filling}

The PN state on the pyrochlore lattice at 1/4-filling is
given by
\begin{equation}
 \Ket{\Psi_{\sigma}}=\prod_{\braket{ijkl}}A^{\dagger}_{ijkl\sigma}
 \prod_{\braket{i'j'k'l'}}A^{\dagger}_{i'j'k'l'\bar{\sigma}}\Ket{0},
\end{equation}
where the sum $\braket{ijkl}$ ($\braket{i'j'k'l'}$) is taken for all
blue (red) tetrahedra of the pyrochlore lattice as indicated in
Fig.~\ref{fig:lattices}.  The parent Hamiltonian for this state is
constructed as follows
\begin{align}
 &h_{ijkl} -
 \tilde{\varepsilon}_{0}=\lambda_{AA}(1-n_{A\uparrow})(1-n_{A\downarrow})
 + \lambda_{BB}n_{B\uparrow}n_{B\downarrow} \nonumber \\
 &+ \lambda_{CC}n_{C\uparrow}n_{C\downarrow}
 + \lambda_{DD}n_{D\uparrow}n_{D\downarrow}  \nonumber \\
 &+ \lambda_{AB}\left[(1-n_{A\uparrow})n_{B\downarrow}
 + n_{B\uparrow}(1-n_{A\downarrow})\right] \nonumber \\
 &+ \lambda_{AC}\left[(1-n_{A\uparrow})n_{C\downarrow}
 + n_{C\uparrow}(1-n_{A\downarrow})\right]  \nonumber \\
 &+ \lambda_{AD}\left[(1-n_{A\uparrow})n_{D\downarrow}
 + n_{D\uparrow}(1-n_{A\downarrow})\right] \nonumber \\
 &+ \lambda_{BC}\left[n_{B\uparrow}n_{C\downarrow}
 + n_{C\uparrow}n_{B\downarrow}\right] + \lambda_{BD}
 \left[n_{B\uparrow}n_{D\downarrow}
 + n_{D\uparrow}n_{B\downarrow}\right] \nonumber \\
 &+ \lambda_{CD}\left[n_{C\uparrow}n_{D\downarrow}
 + n_{D\uparrow}n_{C\downarrow}\right].
\end{align}
We set the parameters assuming time-reversal symmetry as
\begin{align}
 \lambda_{BB}&=\lambda_{CC}=\lambda_{DD}=\lambda_{BC}
 =\lambda_{BD}=\lambda_{CD}, \nonumber \\
 \lambda_{AB}&=\lambda_{AC}=\lambda_{AD},
 \label{setpara2}
\end{align}
then the relations between $\lambda$ and the parameters of the
Hamiltonian are identified as
\begin{align}
\lambda_{AA}&=6t+\frac{1}{2}U-9W, \nonumber \\
\lambda_{BB}&=-2t+\frac{1}{2}U+7W, \nonumber \\
\lambda_{AB}&=-2t-\frac{1}{2}U+9W,
\end{align}
with the relations
\begin{equation}
 X=t-3W,\qquad
  \tilde{\varepsilon}_{0}=\frac{3}{2}(-2t+W).
\end{equation}
The ground-state energy per plaquette is given by
\begin{align}
 \tilde{\varepsilon}_{0}=&-6t-\frac{1}{2}U+9W\nonumber\\
 &+\left(3t+\frac{1}{2}U-\frac{15}{2}W\right)
 \sum_{\sigma}N_{ijkl\sigma}
\end{align}
For bulk systems, it follows from the relation between the number of
plaquettes and the number of sites, $N_{\rm plaq}=(1/2)N_{\rm site}$
that the ground state energy per site becomes
\begin{equation}
 \varepsilon_{0}=\frac{1}{4}U-3W.
\end{equation}
For edged systems, the Hamiltonian with the exact PN states should be
\begin{equation}
 \mathcal{H}_{\rm edge}=\mathcal{H}_{\rm bulk}
  +\left(3t+\frac{1}{2}U-\frac{15}{2}W\right)\sum_{i\in{\rm edge}}n_i
\end{equation}
and its ground-state energy is given by
\begin{equation}
 E_0=\varepsilon_0N_{\rm site}+(6t+U-15W)n_{\rm edge}
\end{equation}
where $n_{\rm edge}$ is the number of the localized electrons at the
 edge.

It follows form the conditions $\lambda\geq0$ and
$\lambda_{AA}+\lambda_{AB}=4t$ that $t>0$.  Thus the exact ground state
is given by the following conditions
\begin{align}
\frac{W}{t} \leq& \frac{2}{3}+\frac{1}{18}\frac{U}{t},\nonumber \\
\frac{W}{t} \geq& \frac{2}{7}-\frac{1}{14}\frac{U}{t},\nonumber \\
\frac{W}{t} \geq& \frac{2}{9}+\frac{1}{18}\frac{U}{t}.
\end{align}
Then we obtain the phase diagrams of this state as shown in
Fig.~\ref{fig:pyrochlore}(a).

\subsection{Plaquette-N\'eel state at 3/4-filling}

The PN state on the pyrochlore lattice at 3/4-filling is
given by
\begin{align}
 \Ket{\Psi_{\sigma}}=&\prod_{\braket{ijkl}}D^{\dagger}_{ijkl\sigma}
  C^{\dagger}_{ijkl\sigma}B^{\dagger}_{ijkl\sigma}\nonumber\\
&\times\prod_{\braket{i'j'k'l'}}
  D^{\dagger}_{i'j'k'l'\bar{\sigma}}
  C^{\dagger}_{i'j'k'l'\bar{\sigma}}B^{\dagger}_{i'j'k'l'\bar{\sigma}}\Ket{0}.
\end{align}
where the sum $\braket{ijkl}$ ($\braket{i'j'k'l'}$) is taken for all
blue (red) tetrahedra of the pyrochlore lattice in
Fig.~\ref{fig:lattices}.  The parent Hamiltonian for this state is
constructed as follows
\begin{align}
 &h_{ijkl}-\tilde{\varepsilon}_{0}
 =\lambda_{AA}n_{A\uparrow}n_{A\downarrow}
 + \lambda_{BB}(1-n_{B\uparrow})
 (1-n_{B\downarrow})\nonumber\\
 &+\lambda_{CC}(1-n_{C\uparrow})(1-n_{C\downarrow})
 +\lambda_{DD}(1-n_{D\uparrow})(1-n_{D\downarrow})
 \nonumber \\
 &+\lambda_{AB}\left[n_{A\uparrow}(1-n_{B\downarrow})
 + (1-n_{B\uparrow})n_{A\downarrow}\right]\nonumber\\
 &+\lambda_{AC}\left[n_{A\uparrow}(1-n_{C\downarrow})
 + (1-n_{C\uparrow})n_{A\downarrow}\right] \nonumber \\
 &+\lambda_{AD}\left[n_{A\uparrow}(1-n_{D\downarrow})
 + (1-n_{D\uparrow})n_{A\downarrow}\right]\nonumber\\
 &+\lambda_{BC}\left[(1-n_{B\uparrow})
 (1-n_{C\downarrow})+(1-n_{C\uparrow})
 (1-n_{B\downarrow})\right] \nonumber \\
 &+\lambda_{BD}\left[(1-n_{B\uparrow})
 (1-n_{D\downarrow})
 +(1-n_{D\uparrow})(1-n_{B\downarrow})\right]\nonumber\\
 &+ \lambda_{CD}\left[(1-n_{C\uparrow})(1-n_{D\downarrow})
 +(1-n_{D\uparrow})(1-n_{C\downarrow})\right].
\end{align}
We set the parameters as Eq.~(\ref{setpara2}) assuming time-reversal
symmetry.  Then the relations between $\lambda$ and the parameters of
the Hamiltonian are identified as
\begin{align}
\lambda_{AA}=&2t+\frac{1}{2}U+15W, \nonumber \\
\lambda_{BB}=&-\frac{2}{3}t+\frac{1}{2}U-W, \nonumber \\
\lambda_{AB}=&-\frac{2}{3}t-\frac{1}{2}U+W,
\end{align}
with the relation,
\begin{equation}
 X=\frac{1}{3}t+W.
\end{equation}
The ground-state energy per plaquette is given by
\begin{align}
 \tilde{\varepsilon}_{0}=&6t-\frac{9}{2}U+9W \nonumber \\
 &+\left(-t+\frac{3}{2}U-\frac{3}{2}W\right)\sum_{\sigma}N_{ijkl\sigma},
\end{align}
where $t$ is negative because of
$15\lambda_{BB}+3\lambda_{AB}=-4t$.
For bulk systems, it follows from the relation between the number of
plaquettes and the number of sites, $N_{\rm plaq}=(1/2)N_{\rm site}$
that the ground state energy per site becomes
\begin{equation}
 \varepsilon_{0}=\frac{9}{4}U.
\end{equation}
For edged systems, the Hamiltonian with the exact PN states should be
\begin{equation}
 \mathcal{H}_{\rm edge}=\mathcal{H}_{\rm bulk}
  +\left(-t+\frac{3}{2}U-\frac{3}{2}W\right)\sum_{i\in{\rm edge}}n_i
\end{equation}
and its ground-state energy is given by
\begin{equation}
 E_0=\varepsilon_0N_{\rm site}+(-2t+3U-3W)n_{\rm edge}
\end{equation}
where $n_{\rm edge}$ is the number of the localized electrons at the
 edge.

Thus the condition of this state is
given as follows
\begin{align}
\frac{W}{|t|} \geq& -\frac{1}{30}\frac{U}{|t|}+\frac{2}{15},\nonumber \\
\frac{W}{|t|}\leq& \frac{1}{2}\frac{U}{|t|}+\frac{2}{3},\nonumber \\
\frac{W}{|t|} \geq& \frac{1}{2}\frac{U}{|t|}-\frac{2}{3}.
\end{align}
Then we obtain the phase diagrams of this state as shown in
Fig.~\ref{fig:pyrochlore}(b).

\begin{figure}[t]
 \includegraphics[width=90mm]{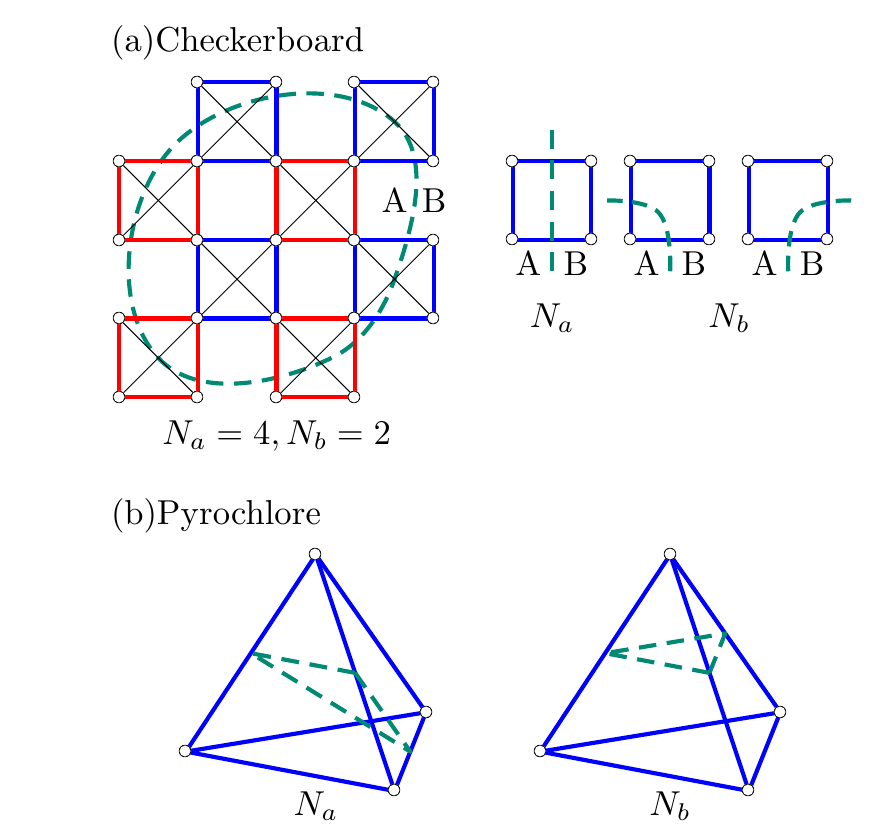}
\caption{Patterns to cut a system into A and B regions to calculate the
EE for (a) the checkerboard and (b) the
Pyrochlore lattice, respectively. $N_a$ and $N_b$ are the numbers of the
cutting lines (surfaces) as indicated.}\label{fig:dividing}
\end{figure}

\section{Entanglement entropy} \label{sec:EE}

In this section we consider the entanglement entropy
(EE)~\cite{Horodecki-H-H-H} of the systems discussed above.  When we
divide the normalized wave function of the system into two regions A and
B as
\begin{equation}
 \ket{\Psi}=\sum_{nm}\Lambda_{nm}
  \ket{\Psi_n^{\rm A}}\otimes\ket{\Psi_m^{\rm B}},
\end{equation}
the EE is given by
\begin{equation}
 S^{\rm A} =-\mathrm{Tr}_{\rm A}
  \left[ \hat{\rho}_{\rm A}\log\hat{\rho}_{\rm A}\right],
\end{equation}
with the reduced density matrix
\begin{equation}
 \hat{\rho}_{\rm A}=\sum_{nm}(\Lambda \Lambda^T)_{nm}
  \ket{\Psi^{\rm A}_n}\bra{\Psi^{\rm A}_m},
  \label{RDM}
\end{equation}
where $\Lambda^T$ is the transposed matrix of $\Lambda$.

The EE for the BN state of the 1D model and that of the PN state in the
Kagom\'e lattice have already been discussed in
Ref.~\onlinecite{Nakamura-N}.  For the 1D model we have
\begin{equation}
 S^{\rm A}=\log 2.
\end{equation}
For the PN state in the Kagom\'e lattice with $1/3$-filling, we get
\begin{equation}
 S^{\rm A}
  =N_{\bigtriangleup}[\log3-(2/3)\log2],
\end{equation}
where $N_{\bigtriangleup}$ means the number of triangles along the
cutting lines. This means that the EE obeys the area law. The EE for the
PN state at $2/3$-filling is obtained as the same value as that of
$1/3$-filling via the particle-hole transformation.

Similarly for the checkerboard lattice, the EE is calculated as
\begin{equation}
 S^{\rm{A}}=N_{a}\log{2}+N_{b}\left(2\log{2}-\frac{3}{4}\log{3}\right),
\end{equation}
where $N_a$ and $N_b$ are the number of two types of cutting lines as
illustrated in Fig.~\ref{fig:dividing}. This result is completely same
as that of the pyrochlore lattice only by changing the definition of the
cutting surface.

\section{Summary} \label{sec:summary}

In summary, we have discussed exact ground states of the generalized
Hubbard model based on the projection operator method.  The Hamiltonian
with the exact ground state can be obtained when the lattices have
bipartite structure in terms of corner sharing unit plaquettes. The
ground states are the plaquette N\'eel states where the spins of the
electrons on the plaquettes form N\'eel order.  We have applied this
method to the checkerboard and pyrochlore lattices, and obtained
parameter regions of the exact ground states for several situations. We
have also obtained exact results for the open systems with localized
electrons at the edges.  Based on the wavefunctions of the exact ground
states, we have calculated the entanglement entropies for the
checkerboard and pyrochlore lattices. The present exact ground states
are the same states used for the discussion of the Berry phases to
detect topological phase transitions by the
multimerization~\cite{Hatsugai-M,Araki-M-H}. Although such multimer
states are often considered in deformed lattice systems without electron
interactions, our results suggest the realization of multimer states in
uniform and correlated electron systems.

\section{ACKNOWLEDGMENTS}

M. N. acknowledges the Visiting Researcher's Program of the Institute
for Solid State Physics, the University of Tokyo.  M.~N. is supported by
JSPS KAKENHI Grant Number 17K05580 and 20K03769.  S.~N. acknowledges
support from the SFB 1143 Project No. A05 of the Deutsche
Forschungsgemeinschaft.



\begin{thebibliography}{99}

 \bibitem{Hubbard}
	 J. Hubbard,
	 \href{https://doi.org/10.1098/rspa.1963.0204}
	 {Proc. R. Soc. London A {\bf 276}, 238 (1963)}.

 \bibitem{Kanamori}
	 J. Kanamori,
	 \href{https://doi.org/10.1143/PTP.30.275}
	 {Prog. Theor. Phys. {\bf 30}, 275 (1963)}.

 \bibitem{Gutzwiller}
	 M. C. Gutzwiller,
	 \href{https://doi.org/10.1103/PhysRevLett.10.159}
	 {Phys. Rev. Lett. {\bf 10}, 159 (1963)}.

 \bibitem{Mott1990}
	 N. F. Mott,
	 Metal-Insulator Transitions (Taylor $\&$ Francis,
	 London/Philadelphia, 1990).

 \bibitem{Tasaki2008}
	 H. Tasaki,
	 \href{https://doi.org/10.1140/epjb/e2008-00113-2}
	 {Eur. Phys. J. B {\bf 64}, 365 (2008)}.

 \bibitem{Fazekas1999}
	 P. Fazekas, Lecture Notes on Electron Correlation and
	 Magnetism (World Scientific, Singapore, 1999).

 \bibitem{Solyom1979}
	 J. S\'olyom,
	 \href{https://doi.org/10.1080/00018737900101375}
	 {Adv. Phys. {\bf 28}, 201 (1979)}.

 \bibitem{Anderson1987}
	 P. W. Anderson,
	 \href{https://doi.org/10.1126/science.235.4793.1196}
	 {Science {\bf 235}, 1196 (1987)}.

 \bibitem{Jaksch2005}
	 D. Jaksch and P. Zoller,
	 \href{https://doi.org/10.1016/j.aop.2004.09.010}
	 {Ann. Phys., {\bf 315}, 52 (2005)}.

 \bibitem{Mazurenko2017}
	 A. Mazurenko, C. S. Chiu, G. Ji, M. F. Parsons,
	 M. Kanm\'{a}sz-Nagy, R. Schmidt, F. Grusdt,
	 E. Demler, D. Greif, and M. Greiner,
	 \href{https://www.nature.com/articles/nature22362}
	 {Nature {\bf 545}, 462 (2017)}.

 \bibitem{Hirsch1}
	 J. E. Hirsch,
	 \href{https://doi.org/10.1103/PhysRevB.40.2354}
	 {Phys. Rev. B {\bf 40}, 2354 (1989)}.

 \bibitem{Hirsch2}
 	 J. E. Hirsch,
	 \href{https://doi.org/10.1103/PhysRevB.40.9061}
	 {Phys. Rev. B {\bf 40}, 9061 (1989)}.

 \bibitem{Hirsch3}
   J. E. Hirsch,
   \href{https://doi.org/10.1103/PhysRevB.43.705}
	 {Phys. Rev. B {\bf 43}, 705 (1991)}.

 \bibitem{Hirsch4}
 	 J. E. Hirsch,
	 \href{https://doi.org/10.1016/0921-4526(90)90194-Y}
	 {Physica B {\bf 163}, 291 (1990)}.

 \bibitem{Campbell-G-L1988}
	 D. K. Campbell, J. T. Gammel, and E. Y. Loh, Jr.,
	 \href{https://doi.org/10.1103/PhysRevB.38.12043}
	 {Phys. Rev. B {\bf 38}, 12043 (1988)}.

 \bibitem{Campbell-G-L1990}
	 D. K. Campbell, J. T. Gammel, and E. Y. Loh, Jr.,
	 \href{https://doi.org/10.1103/PhysRevB.42.475}
	 {Phys. Rev. B {\bf 42}, 475 (1990)}.

 \bibitem{Simon-A}
	 M. E. Sim\'{o}n and A. A. Aligia,
	 \href{https://doi.org/10.1103/PhysRevB.48.7471}
	 {Phys. Rev. B {\bf 48}, 7471 (1993)}.

 \bibitem{Ovchinnikov1993}
	 A. A. Ovchinnikov,
	 \href{https://doi.org/10.1142/S0217984993001442}
	 {Mod. Phys. Lett. B {\bf 7}, 1397 (1993)}.
%
 \bibitem{Strack-V1993}
	 R. Strack and D. Vollhardt,
	 \href{https://doi.org/10.1103/PhysRevLett.70.2637}
	 {Phys. Rev. Lett. {\bf 70}, 2637 (1993)}.

 \bibitem{Strack-V1994}
	 R. Strack and D. Vollhardt,
	 \href{https://doi.org/10.1103/PhysRevLett.72.3425}
	 {Phys. Rev. Lett. {\bf 72}, 3425 (1994)}.

 \bibitem{Arrachea-A}
	 L. Arrachea and A. A. Aligia,
	 \href{https://doi.org/10.1103/PhysRevLett.73.2240}
	 {Phys. Rev. Lett. {\bf 73}, 2240 (1994)}.

 \bibitem{Boer-K-S}
	 J. de Boer, V. E. Korepin, and A. Schadschneider,
	 \href{https://doi.org/10.1103/PhysRevLett.74.789}
	 {Phys. Rev. Lett. {\bf 74}, 789 (1995)}.

 \bibitem{Boer-S}
	 J. de Boer and A. Schadschneider,
	 \href{https://doi.org/10.1103/PhysRevLett.75.4298}
	 {Phys. Rev. Lett. {\bf 75}, 4298 (1995)}.

 \bibitem{Montorsi-C}
	 A. Montorsi and D. K. Campbell,
	 \href{https://doi.org/10.1103/PhysRevB.53.5153}
	 {Phys. Rev. B {\bf 53}, 5153 (1996)}.

 \bibitem{Kollar-S-V}
	 M. Kollar, R. Strack, and D. Vollhardt,
	 \href{https://doi.org/10.1103/PhysRevB.53.9225}
	 {Phys. Rev. B {\bf 53}, 9225 (1996)}.

 \bibitem{Arrachea-A-G}
	 L. Arrachea, A. A. Aligia and E. Gagliano,
	 \href{https://doi.org/10.1103/PhysRevLett.76.4396}
	 {Phys. Rev. Lett. {\bf 76}, 4396 (1996)}.

 \bibitem{Anfossi-D-M}
	 A. Anfossi, F. Dolcini, and A. Montorsi,
	 Recent Research Developments in Physics {\bf 5}, 513
	 (Transworld Research Network, India, 2004),
	 \href{https://arxiv.org/abs/cond-mat/0412532}
	 {cond-mat/0412532}.

 \bibitem{Millan-P-W}
	 J. S. Mill\'an, L. A. P\'erez, and C. Wang,
	 \href{https://doi.org/10.1016/j.physleta.2004.12.080}
	 {Phys. Lett. A {\bf 335}, 505 (2005)}.

 \bibitem{Dobry-A}
	 A. O. Dobry and A. A. Aligia,
	 \href{10.1016/j.nuclphysb.2010.10.017}
	 {Nucl. Phys. B {\bf 843}, 767 (2011)}.

 \bibitem{Majumder-G}
	C. K. Majumder and D. K. Ghosh,
	 \href{https://doi.org/10.1063/1.1664979}
	{J. Math. Phys. {\bf 10}, 1399 (1969)}.

 \bibitem{Affleck-K-L-T}
	I. Affleck, T. Kennedy, E. H. Lieb, and H. Tasaki,
	 \href{https://doi.org/10.1103/PhysRevLett.59.799}
	{Phys. Rev. Lett. {\bf 59}, 799 (1987)}.

 \bibitem{AKLT2}
	I. Affleck, T. Kennedy, E. H. Lieb, and H. Tasaki,
	\href{https://projecteuclid.org/euclid.cmp/1104161001}
	{Commun. Math. Phys. {\bf 115}, 477 (1988)}.

 \bibitem{Itoh}
	 K. Itoh,
	 \href{https://doi.org/10.1143/JPSJ.68.322}
	 {J. Phys. Soc. Jpn. {\bf 68}, 322 (1999)}.

 \bibitem{Itoh-N-M}\label{Itoh-N-M}
	 K. Itoh, M. Nakamura, and N. Muramoto,
	 \href{https://doi.org/10.1143/JPSJ.70.1202}
	 {J. Phys. Soc. Jpn. {\bf 70}, 1202 (2001)}.

 \bibitem{Nakamura-O-I}\label{Nakamura-O-I}
	 M. Nakamura, T. Okano and K. Itoh,
	 \href{https://doi.org/10.1103/PhysRevB.72.115121}
	 {Phys. Rev. B {\bf 72}, 115121 (2005)}.

 \bibitem{Nakamura-I}\label{Nakamura-I}
	 M. Nakamura and K. Itoh,
	 \href{https://doi.org/10.1143/JPSJS.74S.234}
	 {J. Phys. Soc. Jpn. {\bf 74}, 234 (2005)}.

 \bibitem{Nakamura-N}\label{Nakamura-N}
	 M. Nakamura and S. Nishimoto,
	 \href{https://link.springer.com/article/10.1140%2Fepjb%2Fe2018-90352-9}
	 {Eur. Phys. J. B {\bf 91}, 203 (2018)}.

 \bibitem{Hatsugai-M}
	 Y. Hatsugai and I. Maruyama,
	 \href{https://doi.org/10.1209/0295-5075/95/20003}
	 {Europhys. Lett. {\bf 95}, 20003 (2011)}.

 \bibitem{Araki-M-H}
	 H. Araki, T. Mizoguchi, and Y. Hatsugai
	 \href{https://doi.org/10.1209/0295-5075/95/20003}
	 {Phys. Rev. Research {\bf 2}, 012009 (2020)}.

 \bibitem{Horodecki-H-H-H}
	 For example,
	 R. Horodecki, P. Horodecki, M. Horodecki, and K. Horodecki,
	 \href{https://doi.org/10.1103/RevModPhys.81.865}
	 {Rev. Mod. Phys. {\bf 81}, 865 (2009)}.

\end{thebibliography}
\end{document}